\documentclass{article}
\usepackage{spconf,amsmath,graphicx,hyperref}
\usepackage{amssymb}
\usepackage[table]{xcolor} % For cell and row colors
\usepackage{booktabs}      % For better horizontal lines if preferred, though not strictly used here to match the image
\usepackage{threeparttable}

\definecolor{ourmodelgreen}{HTML}{E2EFDA}
\definecolor{secondbestyellow}{HTML}{FEF2CB}

\title{A Lightweight Fourier-based Network for Binaural Speech Enhancement with Spatial Cue Preservation}
\name{
    Xikun Lu$^{1}$ \qquad
    Yujian Ma$^{1}$ \qquad
    Xianquan Jiang$^{2}$ \qquad
    Xuelong Wang$^{3}$ \qquad
    Jinqiu Sang$^{3}$\textsuperscript{*}
        \thanks{* Corresponding author: Jinqiu Sang (jqsang@mail.ecnu.edu.cn)}
}
\address{%
  $^{1}$ Shanghai Institute of Artificial Intelligence for Education, East China Normal University, China\\
  $^{2}$ Boin Hearing Technology (Shanghai) Co., LTD, China\\
  $^{3}$ School of Computer Science and Technology, East China Normal University, China
}
\begin{document}
\ninept
\maketitle
\begin{abstract}
% Binaural speech enhancement presents a trade-off between denoising performance and computational efficiency, particularly with respect to preserving spatial cues. This paper introduces Global Adaptive Fourier Network (GAF-Net), a lightweight deep complex network that achieves an excellent balance of performance and efficiency. Its architecture innovatively integrates a dual-feature auditory encoder, a channel-independent Fourier modulator for efficient global context modeling, and an adaptive gating mechanism to minimize spatial distortion. Experimental results show that GAF-Net achieves state-of-the-art results on binaural cues and MBSTOI while using only a small number of parameters and multiplications. These results establish GAF-Net as a promising solution for high-fidelity binaural audio processing on resource-constrained devices such as hearing aids.

Binaural speech enhancement faces a severe trade-off challenge, where state-of-the-art performance is achieved by computationally intensive architectures, while lightweight solutions often come at the cost of significant performance degradation. To bridge this gap, we propose the Global Adaptive Fourier Network (GAF-Net), a lightweight deep complex network that aims to establish a balance between performance and computational efficiency. The GAF-Net architecture consists of three components. First, a dual-feature encoder combining short-time Fourier transform and gammatone features enhances the robustness of acoustic representation. Second, a channel-independent globally adaptive Fourier modulator efficiently captures long-term temporal dependencies while preserving the spatial cues. Finally, a dynamic gating mechanism is implemented to reduce processing artifacts. Experimental results show that GAF-Net achieves competitive performance, particularly in terms of binaural cues (ILD and IPD error) and objective intelligibility (MBSTOI), with fewer parameters and computational cost. These results confirm that GAF-Net provides a feasible way to achieve high-fidelity binaural processing on resource-constrained devices.

\end{abstract}
\begin{keywords}
Binaural speech enhancement, complex neural network, lightweight model, Fourier network.
\end{keywords}
\section{Introduction}
\label{sec:intro}
Speech enhancement (SE) aims to improve speech quality and intelligibility by reducing background noise~\cite{gannot2017consolidated,zheng2023sixty}. However, for binaural listening devices such as hearing aids, it is essential to enhance speech clarity and preserve the spatial perception encoded by binaural cues~\cite{blauert1997spatial}. While single-channel SE methods excel at noise reduction, their inherent design destroys the critical inter-channel relationships that constitute spatial cues~\cite{leng2021compromise,tokala2023binaural}, making them unsuitable for binaural scenarios. This work focuses on binaural speech enhancement (BSE) to jointly optimize noise suppression and preserve binaural cues.

Statistical signal processing methods are initially applied to BSE tasks, such as the minimum variance distortionless response beamformer~\cite{hadad2015theoretical,as2019robust} and the multichannel Wiener filter~\cite{van2007binaural,van2009speech,tammen2025imposing}. These methods perform well and efficiently under stationary acoustic conditions, but their performance degrades significantly in real-world non-stationary noisy environments. To overcome these limitations, studies have shifted towards deep-learning-based methods, particularly those performing end-to-end modeling in the complex domain~\cite{trabelsi2017deep,hu2020dccrn}, which can generate higher-fidelity speech by jointly optimizing amplitude and phase.

Various deep learning strategies have emerged in the BSE field. Some studies attempt to maximize the signal-to-noise ratio (SNR) by processing binaural channels independently~\cite{tokala2022binaural}, at the expense of spatial fidelity. Other studies introduce complex-valued networks, but their design choices hinder the ability to fully exploit the advantages of complex-domain processing~\cite {sun2019deep}. The demand for higher performance has led to the emergence of more sophisticated architectures, including parallel networks~\cite{han2020real,tokala2023binaural} and complex Transformers~\cite{tokala2024binaural}, with the latter achieving state-of-the-art results. However, their large model size and computational overhead pose significant deployment barriers. To address this, efficient models have been proposed~\cite{wang2025lightweight}, which reduce complexity by selectively processing low-frequency components. 

Despite these advances, existing methods still face several key limitations. First, the prevalent reliance on single short-time Fourier transform (STFT) features, of which the inherent trade-off between time and frequency resolution limits the model's acoustic representation capabilities. Second, a fundamental trade-off exists in effectively modeling long-range temporal dependencies: high-performance architectures like Transformers are computationally prohibitive, while lightweight alternatives often lack sufficient receptive field. 
% Finally, directly selecting low-frequency components for processing can easily introduce over-processing artifacts.

To address these challenges, this paper introduces the \textbf{G}lobal \textbf{A}daptive \textbf{F}ourier \textbf{Net}work (GAF-Net), a novel lightweight complex-valued network designed to achieve a balance among performance and efficiency. The effectiveness of our approach stems from three architectural innovations. First, We propose a dual-feature encoding and fusion module that simultaneously processes STFT-based features and psychoacoustically inspired features to form more robust input representations. Second, we introduce a lightweight Global Adaptive Fourier Modulator (GAFM) as the network backbone, which efficiently models global temporal dynamics in the Fourier domain while maintaining inter-channel independence to preserve spatial cues. Finally, we employ a soft-boundary dynamic gating mechanism to suppress processing artifacts and enhance perceptual naturalness by dynamically blending enhanced and original spectrograms.

\begin{figure*}
    \centering
    \includegraphics[width=1.0\linewidth]{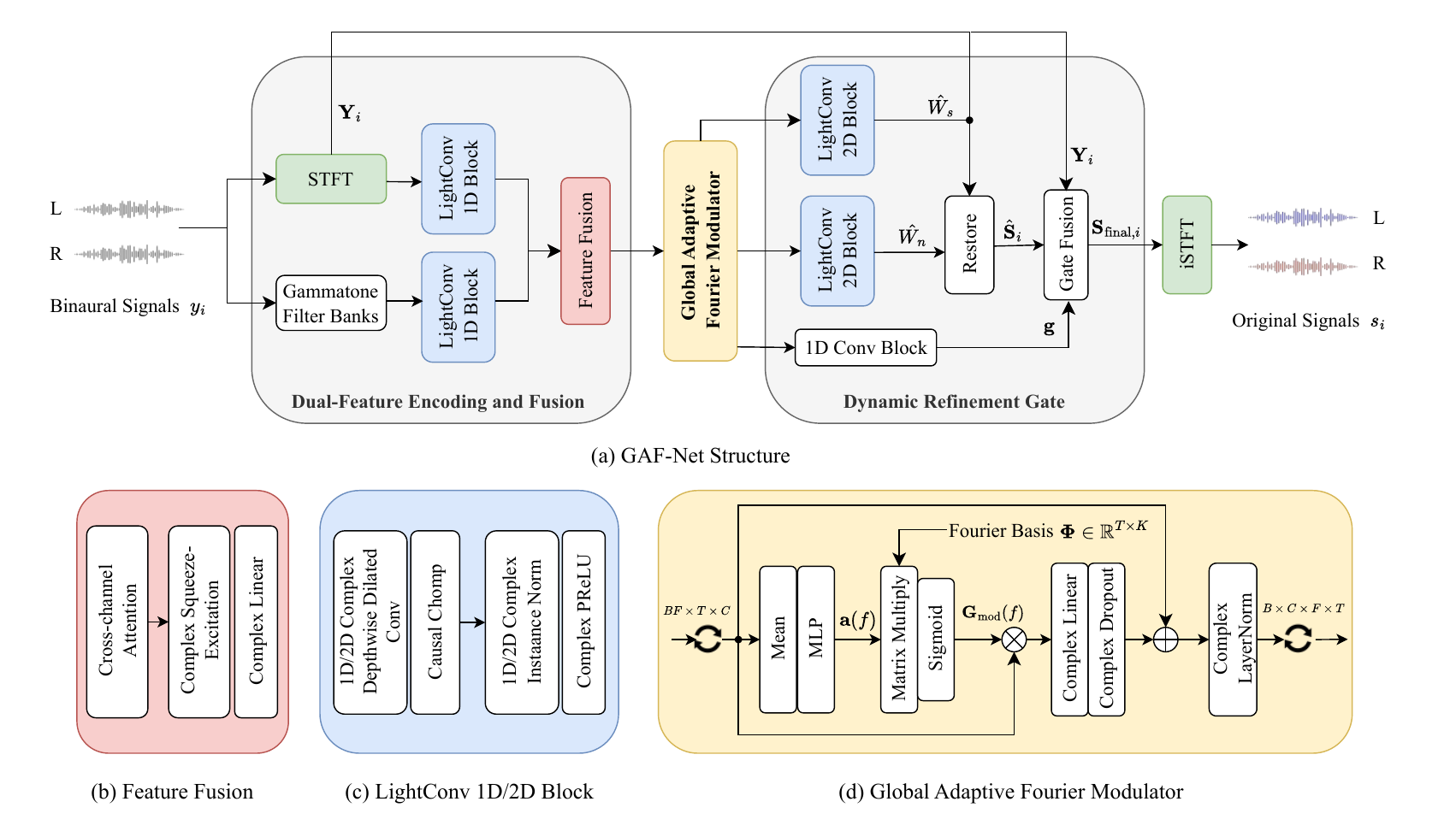}
    \caption{The proposed GAF-Net model, which mainly consists of dual-feature encoding and fusion, global adaptive Fourier modulator, and dynamic refinement gate. Different modules are marked with different colors.}
    \label{fig1}
\end{figure*}

\section{Proposed method}
\label{sec:format}

The objective of BSE is to recover clean speech signals $s_i(t)$ from noisy mixtures $y_i(t) = s_i(t) + n_i(t)$, where $i \in \{L, R\}$, while preserving spatial cues. This task is typically addressed in the time-frequency (T-F) domain, where the relationship is $\mathbf{Y}_i = \mathbf{S}_i + \mathbf{N}_i$. We propose a deep complex network, $\mathcal{G}$, that learns a direct mapping from the noisy to the estimated clean spectrograms: $\mathcal{G}: (\mathbf{Y}_L, \mathbf{Y}_R) \to (\mathbf{S}_L, \mathbf{S}_R)$. As depicted in Fig.~\ref{fig1}, our model employs an encoder-decoder architecture featuring three core components: a dual-feature encoding and fusion module, a GAFM backbone, and a Dynamic Refinement Gate (DRG).

\subsection{Dual-Feature Encoding and Fusion}

To construct a comprehensive acoustic representation, a dual-stream encoder architecture is employed. The primary path generates standard complex STFT spectrograms, whereas the secondary path utilizes a gammatone filterbank to produce a perceptually-motivated representation. The spectrogram features from each path are first processed by separate encoders consisting of $M$ layers of LightConv 1D blocks to extract high-level representations, denoted as $\mathbf{Z}_{\text{stft}}$ and $\mathbf{Z}_{\text{gamma}}$, respectively. These representations are then fused using a cross-channel attention mechanism. Specifically, the magnitude of the gammatone features is used to generate an attention map that modulates the STFT feature path:
\begin{equation}
    \mathbf{Z}_{\text{attended}} = \mathbf{Z}_{\text{stft}} \odot \sigma(\text{Conv}(|\mathbf{Z}_{\text{gamma}}|))
\end{equation}
where $\sigma(\cdot)$ is the sigmoid function. The fused features are further channel-wise recalibrated through a complex Squeeze-and-Excitation (SE) block to generate an overall representation for the backbone network.

\subsection{Global Adaptive Fourier Modulator}

We propose the GAFM as the backbone of our model to efficiently model long-range temporal dependencies. Unlike self-attention mechanisms, which require quadratic complexity, GAFM integrates information by dynamically synthesizing a global filter for each input sequence. 

Let the input to a GAFM block be the complex latent feature $\mathbf{Z}_{\text{bb}} \in \mathbb{C}^{B \times C \times F \times T}$. For each frequency-specific slice $\mathbf{Z}_{f}$, the module first extracts a compact global context vector $\mathbf{c}_f = \mathbb{E}_{T}[|\mathbf{Z}_{f}|]$ by averaging the feature magnitudes along the temporal dimension.
This context vector is then fed into a small multilayer perceptron (MLP) to generate a set of mixing coefficients $\mathbf{a}(f) = \text{MLP}(\mathbf{c}_f) \in \mathbb{R}^{K}$. These coefficients are then used to form a linear combination of a fixed, predefined Fourier basis matrix $\mathbf{\Phi} \in \mathbb{R}^{T \times K}$, thereby synthesizing a content-adaptive gating signal $\mathbf{G}_{\text{mod}}(f)$:
\begin{equation}
    \mathbf{G}_{\text{mod}}(f) = \sigma(\tau \cdot \mathbf{\Phi} \cdot \mathbf{a}(f))
\end{equation}
where $\sigma(\cdot)$ is the sigmoid function. Note that since the synthesis gate $\mathbf{G}_{\text{mod}}(f)$ is a real-valued gate, its element-wise multiplication with the original input  $\mathbf{Z}_{f}$ only modulates the magnitude of the complex features while strictly preserving their phase. This phase-preserving property is crucial for maintaining the Interaural Time Difference (ITD), a key cue for spatial localization~\cite{xie2013head}. Finally, the modulated features are integrated into a standard residual block:
\begin{equation}
    \mathbf{Z}_{\text{out}, f} = \mathbb{C}\text{LN}(\mathbf{Z}_{f} + \mathbb{C}\text{Dropout}(\mathbb{C}\text{Linear}(\mathbf{Z}_{f} \odot \mathbf{G}_{\text{mod}}))
\end{equation}
where $\mathbb{C}\text{Linear}$ denotes a complex linear layer, $\mathbb{C}\text{LN}$ denotes complex layer normalization, and dropout is applied in a way that preserves the complex structure.
By performing these operations in parallel across all frequencies, GAFM efficiently models the global temporal context while ensuring the integrity of spatial information.

\subsection{Binaural Decoding and Dynamic Refinement Gate}

The decoder stage reconstructs the enhanced binaural signals from the backbone's output features $\mathbf{Z}_{\text{out}}$. To this end, two parallel decoder heads are used to estimate Relative Acoustic Transfer Functions (RATFs) \cite{feng2021estimation,feng2022recurrent}. The target speech RATF is denoted as $\hat{\mathbf{W}}_s$, and the noise RATF is denoted as $\hat{\mathbf{W}}_n$. The decoder head consists of $N$ layers of LightConv 2D blocks.

Given the noisy observations and the estimated RATFs, the clean speech spectrograms $\hat{\mathbf{S}}_i$ can be recovered using the following closed-form solution:
\begin{equation} \label{eq:ratf_solve}
\begin{aligned}
    \hat{\mathbf{S}}_{L} = \hat{\mathbf{W}}_s \cdot \hat{\mathbf{S}}_{R}, \quad 
    \hat{\mathbf{S}}_{R} &= \frac{(\mathbf{Y}_L - \hat{\mathbf{W}}_n \mathbf{Y}_R)(\hat{\mathbf{W}}_s - \hat{\mathbf{W}}_n)^*}{(\hat{\mathbf{W}}_s - \hat{\mathbf{W}}_n)^2 + \epsilon} 
\end{aligned}
\end{equation}
where $\epsilon$ is a small constant for numerical stability. 
To suppress potential processing artifacts and enhance model robustness, we introduce the DRG. This module generates a frequency-dependent gate $\mathbf{g} \in [0, 1]^{B\times F}$, which can be understood as a confidence map of the model's enhancement result at each frequency. The gate is derived from the temporal aggregation information of the backbone features:
\begin{equation}
    \mathbf{g} = \sigma(\text{Conv}_{1\times 1}((\text{AvgPool}_T(|\mathbf{Z}_{\text{out}}|)))
\end{equation}

The final output $\mathbf{S}_{\text{final}}$ is obtained by weighted summation of the network-estimated clean spectrogram $\hat{\mathbf{S}}$ and the original noisy spectrogram $\mathbf{Y}$, controlled by $\mathbf{g}$:
\begin{equation}
    \mathbf{S}_{\text{final}, i} = \mathbf{g} \odot \hat{\mathbf{S}}_i + (1 - \mathbf{g}) \odot \mathbf{Y}_i
\end{equation}

This mechanism enables the model to trust its enhancement in high-confidence frequency regions ($\mathbf{g} \to 1$) while restoring to the original signal in the low-confidence areas ($\mathbf{g} \to 0$), thereby achieving a dynamic balance between denoising and fidelity. The refined binaural complex spectrogram is then transformed back to the time domain via the inverse STFT (iSTFT).

\begin{table*}[t]
\centering
\caption{Objective evaluation results under different input SNR conditions. The best results are marked in \textbf{bold}.}
\label{tab:results}

\begingroup
\setlength{\tabcolsep}{4pt}           
\renewcommand{\arraystretch}{1.05}    
\fontsize{8pt}{9.5pt}\selectfont   

\begin{tabular*}{\textwidth}{@{\extracolsep{\fill}} l|cccc|cccc|cccc}
\toprule
\textbf{Input SNR} & \multicolumn{4}{c|}{\textbf{-6 dB}} & \multicolumn{4}{c|}{\textbf{-3 dB}} & \multicolumn{4}{c}{\textbf{0 dB}} \\
\hline
\textbf{Method} & \textbf{MBSTOI $\uparrow$} & \textbf{$\Delta$PESQ $\uparrow$} & \textbf{$\mathcal{L}_{\text{ILD}}$ $\downarrow$} & \textbf{$\mathcal{L}_{\text{IPD}}$ $\downarrow$} &
\textbf{MBSTOI $\uparrow$} & \textbf{$\Delta$PESQ $\uparrow$} & \textbf{$\mathcal{L}_{\text{ILD}}$ $\downarrow$} & \textbf{$\mathcal{L}_{\text{IPD}}$ $\downarrow$} &
\textbf{MBSTOI $\uparrow$} & \textbf{$\Delta$PESQ $\uparrow$} & \textbf{$\mathcal{L}_{\text{ILD}}$ $\downarrow$} & \textbf{$\mathcal{L}_{\text{IPD}}$ $\downarrow$} \\
\hline
DBSEnh~\cite{sun2019deep}          & 0.63 & 0.04 & 7.36 & 1.16 & 0.67 & 0.04 & 7.34 & 1.11 & 0.75 & 0.05 & 6.69 & 1.01 \\
BiTasNet~\cite{han2020real}        & 0.69 & 0.02 & \textbf{4.17} & 1.26 & 0.73 & 0.04 & \textbf{3.98} & 1.16 & 0.75 & 0.03 & \textbf{3.96} & 1.14 \\
BCCTN~\cite{tokala2024binaural}    & 0.71 & 0.08 & 5.91 & 1.08 & 0.78 & 0.14 & 5.60 & 1.02 & 0.80 & \textbf{0.25} & 4.89 & \textbf{0.86} \\
LBCCN~\cite{wang2025lightweight}   & 0.73 & \textbf{0.14} & 7.14 & 1.11 & \textbf{0.80} & \textbf{0.16} & 6.05 & \textbf{1.01} & 0.81 & 0.13 & 6.52 & 1.00 \\
\rowcolor{ourmodelgreen}
\textbf{GAF-Net}                   & \textbf{0.77} & 0.09 & 5.23 & \textbf{0.99} & \textbf{0.80} & 0.14 & 4.89 & 1.02 & \textbf{0.84} & 0.19 & 4.62 & 0.89 \\
\hline
% \hline
\textbf{Input SNR} & \multicolumn{4}{c|}{\textbf{3 dB}} & \multicolumn{4}{c|}{\textbf{6 dB}} & \multicolumn{4}{c}{\textbf{9 dB}} \\
\hline
\textbf{Method} & \textbf{MBSTOI $\uparrow$} & \textbf{$\Delta$PESQ $\uparrow$} & \textbf{$\mathcal{L}_{\text{ILD}}$ $\downarrow$} & \textbf{$\mathcal{L}_{\text{IPD}}$ $\downarrow$} &
\textbf{MBSTOI $\uparrow$} & \textbf{$\Delta$PESQ $\uparrow$} & \textbf{$\mathcal{L}_{\text{ILD}}$ $\downarrow$} & \textbf{$\mathcal{L}_{\text{IPD}}$ $\downarrow$} &
\textbf{MBSTOI $\uparrow$} & \textbf{$\Delta$PESQ $\uparrow$} & \textbf{$\mathcal{L}_{\text{ILD}}$ $\downarrow$} & \textbf{$\mathcal{L}_{\text{IPD}}$ $\downarrow$} \\
\hline
DBSEnh~\cite{sun2019deep}          & 0.80 & 0.05 & 6.41 & 1.00 & 0.84 & 0.10 & 5.94 & 0.89 & 0.84 & 0.12 & 5.56 & 0.83 \\
BiTasNet~\cite{han2020real}        & 0.78 & 0.08 & 3.82 & 1.18 & 0.81 & 0.11 & 3.80 & 1.12 & 0.78 & 0.10 & \textbf{3.67} & 1.13 \\
BCCTN~\cite{tokala2024binaural}    & 0.84 & \textbf{0.32} & 4.40 & \textbf{0.76} & 0.86 & \textbf{0.46} & 4.13 & \textbf{0.67} & \textbf{0.91} & \textbf{0.62} & 3.70 & \textbf{0.57} \\
LBCCN~\cite{wang2025lightweight}   & \textbf{0.85} & 0.23 & 5.80 & 0.96 & \textbf{0.88} & 0.27 & 4.73 & 0.86 & 0.87 & 0.25 & 4.62 & 0.84 \\
\rowcolor{ourmodelgreen}
\textbf{GAF-Net}                   & \textbf{0.85} & 0.27 & \textbf{3.79} & 0.82 & \textbf{0.88} & 0.26 & \textbf{3.46} & 0.71 & 0.88 & 0.24 & 3.32 & 0.66 \\
\hline
% \hline
\textbf{Input SNR} & \multicolumn{4}{c|}{\textbf{12 dB}} & \multicolumn{4}{c|}{\textbf{15 dB}} & \multicolumn{4}{c}{\textbf{Average}} \\
\hline
\textbf{Method} & \textbf{MBSTOI $\uparrow$} & \textbf{$\Delta$PESQ $\uparrow$} & \textbf{$\mathcal{L}_{\text{ILD}}$ $\downarrow$} & \textbf{$\mathcal{L}_{\text{IPD}}$ $\downarrow$} &
\textbf{MBSTOI $\uparrow$} & \textbf{$\Delta$PESQ $\uparrow$} & \textbf{$\mathcal{L}_{\text{ILD}}$ $\downarrow$} & \textbf{$\mathcal{L}_{\text{IPD}}$ $\downarrow$} &
\textbf{MBSTOI $\uparrow$} & \textbf{$\Delta$PESQ $\uparrow$} & \textbf{$\mathcal{L}_{\text{ILD}}$ $\downarrow$} & \textbf{$\mathcal{L}_{\text{IPD}}$ $\downarrow$} \\
\hline
DBSEnh~\cite{sun2019deep}          & 0.87 & 0.07 & 5.38 & 0.80 & 0.89 & 0.08 & 4.79 & 0.71 & 0.78 & 0.06 & 6.18 & 0.93 \\
BiTasNet~\cite{han2020real}        & 0.82 & 0.30 & 3.74 & 1.10 & 0.82 & 0.52 & 3.69 & 1.10 & 0.77 & 0.13 & \textbf{3.86} & 1.17 \\
BCCTN~\cite{tokala2024binaural}    & \textbf{0.92} & \textbf{0.60} & 3.63 & 0.56 & 0.90 & \textbf{0.58} & 3.35 & 0.50 & 0.84 & \textbf{0.35} & 4.59 & 0.79 \\
LBCCN~\cite{wang2025lightweight}   & 0.89 & 0.24 & 4.35 & 0.78 & \textbf{0.92} & 0.30 & 3.83 & 0.72 & 0.85 & 0.20 & 5.32 & 0.88 \\
\rowcolor{ourmodelgreen}
\textbf{GAF-Net}                   & 0.89 & 0.27 & \textbf{3.00} & \textbf{0.50} & \textbf{0.94} & 0.27 & \textbf{2.53} & \textbf{0.44} & \textbf{0.86} & 0.22 & \textbf{3.86} & \textbf{0.75} \\
\bottomrule
\end{tabular*}
\endgroup

\end{table*}

\subsection{Loss Function}

The proposed network is trained by optimizing a composite loss function $\mathcal{L}_{\text{total}}$, which consists of a primary task-oriented loss $\mathcal{L}_{\text{task}}$, and an regularization term $\mathcal{L}_{\text{reg}}$:
\begin{equation}
    \mathcal{L}_{\text{total}} = \mathcal{L}_{\text{task}} + \mathcal{L}_{\text{reg}}
\end{equation}

The primary task loss function $\mathcal{L}_{\text{task}}$ follows the design in~\cite{tokala2024binaural} and is a weighted sum of four objectives, aiming to synergistically optimize denoising performance, speech intelligibility, and the preservation of binaural cues:
\begin{equation}
    \mathcal{L}_{\text{task}} = \alpha\mathcal{L}_{\text{SNR}} + \beta\mathcal{L}_{\text{STOI}} + \gamma\mathcal{L}_{\text{ILD}} + \kappa\mathcal{L}_{\text{IPD}}
\end{equation}

The time-domain terms $\mathcal{L}_{\text{SNR}}$ and $\mathcal{L}_{\text{STOI}}$ represent noise suppression and intelligibility. The time-frequency terms, $\mathcal{L}_{\text{ILD}}$ and $\mathcal{L}_{\text{IPD}}$, ensure accurate reconstruction of spatial information by penalizing the mean absolute error between the Interaural Level Difference (ILD) and Interaural Phase Difference (IPD) of the clean and enhanced signals.

The regularization term $\mathcal{L}_{\text{reg}}$ is applied to the gating signal $\mathbf{g}$ of the DRG to provide fine-grained control over its behavior. It consists of three parts:
\begin{equation}
    \mathcal{L}_{\text{reg}} = \lambda_s \mathcal{R}_{\text{sparse}}(\mathbf{g}) + \lambda_e \mathcal{R}_{\text{entropy}}(\mathbf{g}) + \lambda_{tv} \mathcal{R}_{\text{TV}}(\mathbf{g})
\end{equation}

The three parts are: 1) an L1 sparse regularizer ($\mathcal{R}_{\text{sparse}}$) to promote conservative augmentation strategy; 2) a negative entropy regularizer ($\mathcal{R}_{\text{entropy}}$) to encourage decisive, binary-like gating decisions; and 3) a Total Variation (TV) regularizer ($\mathcal{R}_{\text{TV}}$) to enforce spectral smoothness. 
% This regularization framework enhances model robustness and helps to suppress artifacts.

\section{Experiments}
\label{sec:pagestyle}

\subsection{Datasets}

We construct a binaural speech dataset for training and testing our model. Clean monaural speech signals are sourced from the VCTK corpus~\cite{yamagishi2019cstr}, from which we generate 40,000 training, 5,000 validation, and 5,000 test samples, each with a duration of 2 seconds.
To ensure a robust evaluation of generalization, a strict partitioning of data resources is enforced. Both the speakers from the VCTK corpus and the Head-Related Impulse Responses (HRIRs) from the HUTUBS database~\cite{brinkmann2019cross} are split into training, validation, and test-exclusive subsets. This speaker- and HRIR-disjoint strategy guarantees that the model encounters neither familiar speakers nor familiar acoustic transfer functions during validation and testing.
During data synthesis, each monaural speech utterance is spatialized into a binaural signal by convolving it with a pair of HRIRs randomly selected from the corresponding subset. The target speech source is placed at a random azimuth in the horizontal plane, from -90\textdegree{} to +90\textdegree{}. Noise signals are drawn from the NOISEX-92 database~\cite{varga1993assessment} and are synthesized into a diffuse isotropic noise field by convolving non-overlapping segments of a noise source with all available HRIRs on the horizontal plane for a given subject and summing the results.
For the training and validation sets, noisy mixtures are generated at a random SNR continuously drawn from -7~dB to 16~dB. For the test set, fixed SNR levels ranging from -6~dB to 15~dB in 3~dB steps are used to facilitate a balanced performance assessment. The noise types used for training include white, pink, factory, and babble noise. During evaluation, an additional unseen noise type (car engine noise) is introduced to assess the model's noise generalization capability. All audio data are generated at a sampling rate of 16~kHz.

\subsection{Implementation Details}

All audio signals are processed at a sampling rate of 16~kHz. For feature extraction, we employ an STFT with a 256-point FFT and a 128-sample hop size. The parallel Gammatone filterbank is configured with 64 channels. The model's encoder consists of two layers of LightConv 1D blocks ($M=2$), the backbone network is one layer of GAFM, and the decoder is two layers of LightConv 2D blocks ($N=2$). 
For training, the AdamW optimizer is used with an initial learning rate of $2 \times 10^{-4}$. The model is trained for 100 epochs with a batch size of 20, using a multi-step learning rate scheduler. Early stopping is enabled to terminate training if the validation loss does not improve for 8 consecutive epochs. The weighting coefficients $\alpha, \beta, \gamma,$ and $\kappa$ for the composite loss function are set to 1, 10, 1, and 10, respectively. The regularization weight $\lambda_s$, $\lambda_e$, and $\lambda_{tv}$ are all set to $1 \times 10^{-4}$. Further implementation details and full parameter settings can be found in our open-source code.\footnote{\url{https://github.com/Luxikun669/GAF-Net}}

\subsection{Evaluation Metrics}

To comprehensively evaluate model performance, four standard objective metrics are employed. The gain of perceptual evaluation of speech quality ($\Delta$PESQ)~\cite{rix2001perceptual} quantifies the model's perceptual quality improvement relative to the noisy input. The ILD error ($\mathcal{L}_{\text{ILD}}$)~\cite{tokala2024binaural} and IPD error ($\mathcal{L}_{\text{IPD}}$)~\cite{tokala2024binaural} are used to measure the reconstruction accuracy of spatial cues such as sound source localization. Finally, the Binaural Short-Time Objective Intelligibility (MBSTOI)~\cite{andersen2018refinement} serves as a comprehensive metric that jointly assesses speech intelligibility and the integrity of binaural cues.

% Table for Computational Complexity
\begin{table}[t]
\centering
\caption{Parameter count and computational complexity.}
\label{tab:complexity}
\begingroup
\setlength{\tabcolsep}{3.5pt}
\renewcommand{\arraystretch}{1.05}
\fontsize{9pt}{10pt}\selectfont
% \begin{tabular}{l|ccc}  @{\extracolsep{\fill}}
\begin{tabular}{l|@{\hspace{10pt}}c@{\hspace{16pt}}c@{\hspace{16pt}}c}
\toprule
\textbf{Method} & \textbf{\# Param. $\downarrow$} & \textbf{MACs $\downarrow$} & \textbf{RTF $\downarrow$} \\
\midrule
DBSEnh     & 10.5 M   & 0.99 G & \textbf{0.026} \\
BiTasNet   & 1.7 M    & 4.97 G & 0.148          \\
BCCTN      & 11.1 M   & 16.38 G & 0.237          \\
LBCCN      & \textbf{38.0 K} & \textbf{0.30 G} & 0.092 \\
\textbf{GAF-Net} & 129.0 K & 2.79 G & 0.150 \\
\bottomrule
\end{tabular}
\endgroup
\end{table}

\begin{table}[t]
\centering
\caption{Ablation study of the proposed model components.}
\label{tab:ablation}
\begingroup
\setlength{\tabcolsep}{3.2pt}
\renewcommand{\arraystretch}{1.05}
\fontsize{9pt}{10pt}\selectfont
\begin{threeparttable}
\begin{tabular*}{\columnwidth}{@{\extracolsep{\fill}} l|cccc}
\toprule
\textbf{Method} & \textbf{MBSTOI $\uparrow$}  & \textbf{$\Delta$PESQ $\uparrow$} & \textbf{$\mathcal{L}_{\text{ILD}}$} $\downarrow$ & \textbf{$\mathcal{L}_{\text{IPD}}$ $\downarrow$} \\
\midrule
\textbf{GAF-Net}    & \textbf{0.86} & 0.22  & \textbf{3.86} & \textbf{0.75} \\
w/o Gammatone       & 0.81 & 0.11  & 5.10 & 0.77 \\
w/o GAFM            & 0.83 & 0.20  & 4.99 & 0.80 \\
w/o DRG             & 0.85 & \textbf{0.31}  & 4.61 & 1.00 \\
Global DRG\textsuperscript{a} & 0.85 & 0.19  & 4.73 & 0.76 \\
% w/o $\mathcal{L}_{\text{reg}}$ & 0.84 & 0.26 & 4.66 & 4.13 & 0.77 \\
\bottomrule
\end{tabular*}%
\vspace{2pt}
\begin{tablenotes}[flushleft]  %  [para,flushleft]
\normalsize
\item[a] Global DRG generates a fixed gating factor $\mathbf{g}$ for each frequency.
\end{tablenotes}
\end{threeparttable}
\endgroup
\end{table}

\section{Results and discussion}
\label{sec:results}

We evaluate the performance of the proposed GAF-Net and compare it with several baseline models~\cite{sun2019deep,han2020real,tokala2024binaural,wang2025lightweight}. As shown in Table~\ref{tab:results}, GAF-Net performs well on key metrics that impact the binaural listening experience. The GAF-Net model's strength lies in its accurate reconstruction of binaural intelligibility and spatial cues. GAF-Net achieves the highest average MBSTOI score, as well as the lowest average $\mathcal{L}_{\text{IPD}}$ and $\mathcal{L}_{\text{ILD}}$, outperforming all baseline models. Furthermore, $\Delta$PESQ result is suboptimal. Across speech quality metrics, the results reveal a profound trade-off. While large models like BCCTN demonstrate superior $\Delta$PESQ, this often comes at the expense of higher spatial distortion and significant computational overhead. In contrast, GAF-Net maintains strong competitiveness on these metrics while also leading the way in binaural fidelity.

Furthermore, we found that GAF-Net achieves a remarkable balance between performance and efficiency, as shown in Table~\ref{tab:complexity}. GAF-Net contains only 129.0 K parameters and 2.79 GMACs, with a computational cost of only 1.2\% of larger models like BCCTN, yet surpasses them on key binaural performance metrics. Compared to LBCCN, GAF-Net achieves significant performance improvements with a slight increase in resources. Its real-time factor (RTF)\footnote{The RTF is measured with Intel(R) Xeon(R) Gold 6146 CPU.} demonstrates its suitability for deployment on resource-constrained devices. Therefore, GAF-Net stands out as a computationally efficient and high-performance solution, successfully bridging the performance and efficiency gap between existing methods.

The ablation study in Table~\ref{tab:ablation} deconstructs the source of the GAF-Net model's performance advantage. Removing features such as gammatone features or the GAFM backbone leads to an overall drop in performance, confirming the importance of dual-feature representation and global context modeling. Notably, analysis of DRG reveals a key trade-off: removing DRG improves $\Delta$PESQ, but significantly worsens $\mathcal{L}_{\text{IPD}}$. This suggests that DRG intelligently trades off a degree of aggressive denoising in exchange for a significant reduction in spatial distortion and audible artifacts, which is crucial for a high-fidelity binaural listening experience.

\section{Conclusion}
This paper introduced GAF-Net, a lightweight deep complex network designed to address the performance-efficiency trade-off in binaural speech enhancement. With its innovative architecture, GAF-Net efficiently models global context while preserving spatial cues. With only 129 K parameters, it achieves SOTA results on key metrics such as spatial completeness ($\mathcal{L}_{\text{IPD}}$ and $\mathcal{L}_{\text{ILD}}$) and binaural clarity (MBSTOI). It also maintains strong competitiveness in speech quality compared to larger models. These findings validate GAF-Net as an efficient, low-resource solution for high-fidelity binaural audio applications. However, the current study is primarily limited to controlled anechoic environments to validate specific binaural cue preservation. Future work will focus on evaluating its robustness in reverberant environments and extending its core components to other multichannel audio tasks.

\section{ACKNOWLEDGEMENTS}
This work was supported by Shanghai Municipal Commission of Economy and Informatization (No. 2024-GZL-RGZN-02001).

\bibliographystyle{IEEEbib}
\bibliography{strings,refs}

\end{document}